\documentclass[12pt]{article}
\pdfoutput=1
\usepackage{fullpage}
\usepackage{amsmath}
\usepackage{lineno}
\usepackage{url}
\usepackage{xcolor}

\usepackage{graphicx}

\title{\bf The tipping effect of delayed interventions on the evolution of COVID-19 incidence}

\author{ \small Kristoffer Rypdal {\small Department of Mathematics and Statistics} \\ {\small UiT -- The Arctic University of Norway}}

\date{}

\begin{document}
\maketitle

\begin{abstract} We combine infectious disease transmission and the non-pharmaceutical intervention (NPI) response to disease incidence into one closed model consisting of two coupled delay differential equations for the incidence rate  and the time-dependent reproduction number. The model contains three free parameters, the initial reproduction number, the intervention strength, and the response delay relative to the time of infection. The NPI response is modeled by assuming that the rate of change of the reproduction number is proportional to the negative deviation of the incidence rate from an intervention threshold. This delay dynamical system exhibits damped oscillations in one part of the parameter space, and growing oscillations in another, and these are separated by a surface where the solution is a strictly periodic nonlinear oscillation. For parameters relevant for the COVID-19 pandemic, the tipping transition from damped to growing oscillations occurs for response delays of the order of one week, and suggests that effective control and mitigation of successive epidemic waves cannot be achieved unless NPIs are implemented in a precautionary manner, rather than merely as a response to the present incidence rate.
\end{abstract}

\section{Introduction}
A year after the COVID-19 pandemic began its rapid geographic expansion across the globe, it is evident that the incidence rate in each country evolves in waves.  In  Europe, the typical pattern so far has been two waves, the first in the spring of 2020, and a second longer and stronger wave that started in the fall and is still ongoing at the time of writing, February 2021 \cite{ECDC}. The increased strength of the second wave is particularly prominent in the case notification rate, but part of this increase is due to increasing testing rate. Nevertheless, the tendency is clear also in the  reported COVID-19 death rates, although many countries have seen lower death rates in the start of the second wave, because this wave began with infection spreading in the younger age groups. 

Based on the experience from the first wave, countries should have been more prepared for the second wave than for the first, and  mitigation should have been be more feasible. There is no apparent microbiological mechanisms that could have driven the strong second wave, even though new and more contagious mutants have started to make an impact as we    enter the calendar year of 2021. Hence,  the explanation seems to be associated to how our societies respond to the threats of the pandemic.  

Some evidence suggest that the wavy pattern during the first year of the pandemic has been driven by an interaction between the pathogen' natural tendency to reproduce and the non-pharmaceutical interventions (NPIs) implemented by governments \cite{Rypdal2020}. Even though the emergence of new and more contagious mutants of the SARS-CoV-2 virus with higher reproduction numbers ${\cal R}$  and the roll-out of vaccines will play an increasingly important role in reducing ${\cal R}$,  NPIs are expected to play an important role in regulating and controlling the incidence rate also in the upcoming year.

In this paper the term incidence rate $X(t)$ will refer to the daily number of actual infections taking place in a country at the time $t$. It does not refer to the recorded incidence (case notification rate), and the time $t$ is the time of infection, not the time the infection is detected. The instantaneous reproduction number ${\cal R}(t)$ refers to the average number of new infections transmitted by an infected individual at the time $t$.  This means that these quantities are not directly  observable at the time $t$.

 There is clear and strong correlation between case notification rate and NPIs in most countries, and the time lag between NPIs and changes in recorded incidence corresponds roughly
 to the sum of incubation period and time for testing, analysis and registration. Thus, it is reasonable to assume that the effect of NPI-induced changes of  ${\cal R}(t)$ on the actual transmission of the infection is more or less instantaneous. The same is not the case with the effect of disease incidence on the NPIs. Here we would expect considerable delay between cause and effect, a delay  we shall refer to as the social response time.
 
NPIs represent a great burden on society, and so far in the COVID-19 pandemic there are very few examples where interventions have been effectuated in a precautionary manner.  Political pressure has forced policy makers to respond to the  {\em recorded} disease burden, which is delayed by 1-2 weeks relative to the actual state,  even though most governments have access to model projections that can inform them about the true present state of the epidemic and the likely development in the near future. The objective of this paper is to investigate whether or not this delay may have an important influence on the trajectory of the epidemic state.

It is intuitively evident that NPIs effectuated as responses to the true epidemic state will lead to oscillations in the disease incidence. This is because NPIs act as a restoring force counteracting the virus' natural tendency to reproduce, while the disease activity level below or above a socially acceptable threshold will enhance or reduce the NPIs. In a recent paper \cite{Rypdal2020}, we constructed a simple model that reduces to a damped harmonic oscillator in the small-amplitude (linearized) limit. In that paper  we demonstrated that a weakening of the intervention response over time could counteract the damping and lead to stronger and   longer secondary waves, but it was assumed that the intervention response is instantaneous. In the present paper, we explore a similar model, where the intervention fatigue is replaced by a delayed response.

In Section \ref{sec:Methods} we formulate and explain the mathematical model, which takes the form of a system of first-order delay differential equations \cite{DDEbook}, and we discuss briefly the nature of the equilibria and a possible limit cycle of the system and their relation to three model parameters expressing the reproductive ability of the pathogen, the intervention strength, and the response delay. Then we explore the solutions of the system numerically in Section \ref{sec:results}, where  we demonstrate the existence of a tipping transition that transforms the solution  from  damped  into growing oscillations, and we map the surface in the parameter space where this transition takes place. In  Section \ref{sec:discuss} we discuss the possible policy implications  of these results.
\section{Methods}
\label{sec:Methods}
\subsection{Evolution of epidemic state under given social evolution}
\label{sec:epidemic evolution}
Let $J(t)$ be the cumulative fraction of infected individuals in a population, and $I(t)$ the instantaneous fraction of infectious individuals. The time evolution of these quantities can be modelled by the simple system of ordinary differential equations,
\begin{linenomath*}
\begin{eqnarray} 
d_tJ(t)&=&\alpha {\cal R}(t) I(t),   \label{1}\\ 
\label{2} 
dI_t(t)&=&\alpha[{\cal R}(t) -1] I(t)\,. 
\end{eqnarray}
\end{linenomath*}
Here the notation $d_t$ stands for the derivative with respect to time, ${\cal R}(t)$ is the effective reproduction number at time $t$ and $\alpha^{-1}$ is the mean duration of the infectious period. The system is a reformulation of the standard Susceptible-Infectious-Recovered (SIR) model of Kermack and McKendrick \cite{SIR}. Here, the reproduction number can be written in the form,
\begin{linenomath*}
\begin{equation}\label{3}
{\cal R}(t)=\frac{\beta(t)}{\alpha}S(t),
\end{equation}
\end{linenomath*}
where  $S(t)=1-J(t)$ is the fraction of susceptible individuals in the population  and $\beta(t)$ is the contact rate, i.e.,  $\beta(t)^{-1}$ is the characteristic time between contacts. The important point is that $\alpha$ is time-independent and determined by the pathogen, $\beta(t)$ is completely determined by the contagiousness of the pathogen and the evolution of the social state, while $S(t)$ depends on the degree of immunity in the population. In the remainder of this paper we shall assume that the degree of herd immunity does not change significantly during the time span of the study, implying that we can consider $S\approx S_0$ as approximately constant, and hence that ${\cal R}(t)\approx (S_0/\alpha) \beta(t)$ only varies in time due to variations in the social conditions that determine the contact rate $\beta(t)$.
Note also that $\alpha [{\cal R}(t)-1]$ is the relative growth rate for the infectious fraction, $\gamma_I(t)\equiv d_t\ln{I(t)}$, which is positive when ${\cal R}(t)>1$ and negative when ${\cal R}(t)<1$.

\subsection{Evolution of social state under given epidemic evolution}
\label{sec:social evolution}
 Eqs.\;(\ref{1}--\ref{3}) describe the dynamics of the epidemic state $J(t)$ and $I(t)$ when the evolution of the  social state represented by ${\cal R}(t)$ is given.  A closed model can only be obtained by adding a description of the  response of the social contact rate to the epidemic state. We shall represent this response by assuming that the relative rate of change  $\gamma_{\cal R}\equiv d_t\ln{\cal R}(t)$ is a  linear function of the delayed incidence rate $X(t-d)\equiv d_tJ(t-t_d)$, where $t_d$ is the time delay. This function is positive when $X(t-t_d)$ is below a threshold $X^*$ and negative when it is above that threshold. Society reacts to the incidence rate only when it receives the information about new infections, which is the reason for the delay. When the incidence rate is low, society responds by relaxing restrictions, and the reproduction number increases. When the incidence rate exceeds the threshold $X^*$, restrictions are introduced that make $d_t \ln{\cal R}(t)$ to change sign from positive to negative. Thus, we end up with the equation,
\begin{linenomath*}
\begin{equation}
 d_t\ln{\cal R}(t)= -k [X(t-t_d)-X^*],   \label{4}
\end{equation} 
\end{linenomath*}
where $k$ is a coefficient which characterizes the strength of the social response to the epidemic evolution, we shall refer to it as the intervention strength parameter.

\subsection{A closed model for the socio-epidemic state}
In the following, it is convenient to introduce a dimensionless time variable $t \rightarrow \alpha t$, which allows us to formulate the differential equations as functions of time measured in units of the infectious time $\alpha^{-1}$, rather than days.  We also express the incidence rate and the infectious fraction in units of the intervention threshold, i.e., we introduce the dimensionless variables $X(t)\rightarrow X(t)/X^*$ and $I(t)\rightarrow I(t)/X^*$.  Eqs.\;(\ref{1}) can then be written as $X(t)={\cal R}(t)I(t)$ and inserted into Eq. (\ref{2}), which leaves us with the following nonlinear system of delay differential equations;

\begin{linenomath*}
\begin{eqnarray}
d_t{\cal R}(t)&=& -\kappa [{\cal R}(t-\delta)I(t-\delta)-1]{\cal R}(t), \label{5}\\ 
d_tI(t)&=& [{\cal R}(t)-1]I(t), \label{6}
\end{eqnarray}
\end{linenomath*}
where $\kappa=kX^*/\alpha $ and $\delta=t_d\alpha$. Since this is a system of delay differential equations we have to specify the state variables in the time interval $t\in (-\delta,0)$ rather than only at the time $t=0$ as in a conventional initial value problem. For this particular problem, we can do this in a way that reflects the actual epidemiological situation. In the early stage of the epidemic, the reproduction number is ${\cal R}_0$ which is determined by the infectivity of the pathogen and the social structure in the actual country in absence of any non-pharmaceutical interventions. Let us define the time origin $t=0$ as the time when interventions start. In  Eq. (\ref{5}) the threshold for intervention is $X=1$, but because of the delay, the intervention that starts to change ${\cal R}$ at $t=0$ is a response to the reported incidence rate which took place at $t=-\delta$, which means that $X(-\delta)=1$. Since ${\cal R}(t)={\cal R}_0$ for $t\in(-\delta,0)$ we have that $X(-\delta)={\cal R}_0I(-\delta)=1$, i.e., $I(-\delta)=1/{\cal R}_0$. Eq. (\ref{6}) is valid not only for $t>0$, but also in the time interval $(-\delta,0)$ when ${\cal R}(t)={\cal R}_0$, so the solution for $I(t)$ satisfying the condition $I(-\delta)=1/{\cal R}_0$ in this interval yields the following ``initial conditions'' for the interval $t\in (-\delta,0)$;
\begin{linenomath*}
\begin{equation}
{\cal R}(t)={\cal R}_0 ,\;\;I(t)=(1/{\cal R}_0)\exp{[({\cal R}_0-1)(t+\delta)]}. \label{7}
\end{equation}
\end{linenomath*}
Note also, that this choice does not only make epidemiological sense, but also ensures continuity in the derivatives of ${\cal R}(t)$ and $I(t)$ across the intervention point $t=0$.

The system has two equilibrium states: a fixed point in ${\cal R}=1$ and  $I =1$, where the number of infected stays constant at the threshold value, and another fixed point in ${\cal R}=0$ and  $I =0$, which is a state with no transmission and nobody infected. The latter is obviously a repellor: if ${\cal R}(t)$ and $I(t)$ both are becoming very small, then Eq. (\ref{6}) implies that $I(t)$ decays exponentially towards zero as $I(t)\approx \exp (-t)$, while Eq. (\ref{5}) implies that ${\cal R}(t) \approx \exp (\kappa t)$ grows exponentially. The equilibrium $({\cal R},I)=(1,1)$ a stable spiral node for some regions of the parameter space $({\cal R}_0,\kappa,\delta)$, it is an unstable spiral node in another region, and these regions are separated by a surface in parameter space where the solution is a limit cycle. 

\subsection{Numerical exploration of the parameter space}
Delay differential equations are integrated numerically by the same methods as ordinary differential equations, and these fast routines can be used to explore the nature of the solutions in the regions of the parameter space which is of interest to the COVID-19 pandemic. In particular, the region close to the transition surface is carefully mapped and we can easily detect the transition points with three decimal accuracy in those three parameters. In practice, we run the routine for an array of  values of the parameters ${\cal R}_0$ and $\kappa$ and detect the value of $\delta$ for which the solution shifts from a decaying oscillation to a growing one. It is observed that this transition is sharper for larger values of ${\cal R}_0$ and smaller values of $\kappa$. This is reasonable, since this is the region of the parameter space corresponding to stronger initial epidemic spread and weaker interventions.

\section{Results}
\label{sec:results}
The delay-differential equation system presented in Eqs. (\ref{5})-(\ref{7}) is an interesting object that warrants further mathematical study. The purpose of this paper, however, is not
a mathematical exploration, but to extract those properties of the system that are of particular relevance to the delayed social response to changing reported incidence of the infection with the SARS-CoV-2 virus.

 The introduction of dimensionless, independent and dependent variables has revealed that the epidemic evolution depends on the response time delay $\delta$ measured in units of the effective infectious time $\alpha^{-1}$. For COVID-19,  a reasonable estimate of $\alpha^{-1}$ is 8 days \cite{Rypdal2020}, which means that the physical delay in units of days is given by $t_d= \delta/\alpha=8\,\delta$ days. Hence, it  seems reasonable to explore delay times  in a range around $\delta \sim 1$.  

The interpretation of the dimensionless intervention strength $\kappa$  can be seen from Eq. (\ref{5}) if we consider a state with very low incidence $I(t)\ll 1$, such that the first term on the right hand side can be neglected. In that case, we can write $\kappa \approx d_t\ln {\cal R}$, which is the relative rate of change of ${\cal R}(t)$ measured in time units of 8 days.  From experience with the first wave of the COVID-19 epidemic, we have seen that a characteristic time scale of change of ${\cal R}$ varies from a few weeks to a few months, which is included in the $\kappa$-range we  explore below; $\kappa\in (0.2,1.0)$.

\begin{figure}[t!]
    \centering
\includegraphics[width=14 cm]{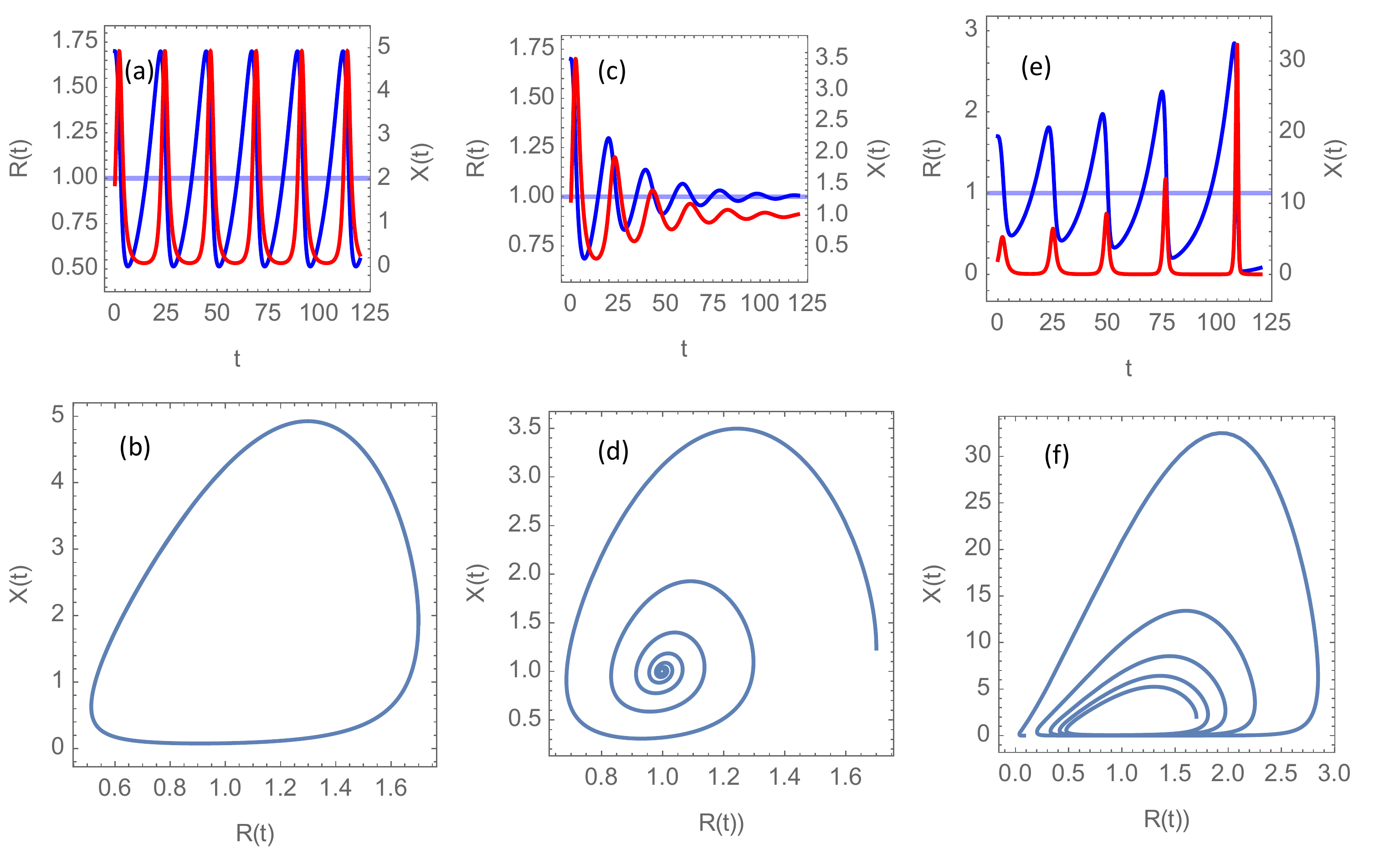}
    \caption{\footnotesize Solutions of the delay differential system Eqs. (\ref{5}- \ref{7}) for initial reproduction number ${\cal R}_0=1.7$ and normalized intervention strength $\kappa=0.1$. Blue curves show the evolution of the time-dependent reproduction number ${\cal R}(t)$ and red curves the incidence rate $X(t)$ in units of the intervention threshold.  Panels (a), (c), and (e) show the graphs for ${\cal R}(t)$ and $X(t)$  for delay times for intervention $\delta=0.90272, 0.3,$ and 1.0, respectively. Times are given in units of infection duration $\alpha^{-1}$. Panels (b), (d), and (f) present the corresponding phase portraits, i.e, the trajectories for the vector (${\cal R}(t),X(t)$) as a parameterized curve for the same values of $\delta$.}
    \label{fig:1}
\end{figure}

The three characteristic modes of epidemic development are shown in Figure \ref{fig:1}. On a ``transition  surface'' in the $({\cal R}_0,\kappa,\delta)$ parameter space, the solution to the system Eqs. (\ref{5})-(\ref{7}) is a nonlinear oscillation (limit cycle), as shown in Figure \ref{fig:1}(a,b), but this oscillation is parametrically unstable. This means that an infinitesimal perturbation of the parameters could lead  either to a damped oscillation, as in Figure \ref{fig:1}(c,d), or to a growing oscillation,  as in Figure \ref{fig:1}(e,f), depending on which side of the transition surface the perturbed parameter point is located. 

\begin{figure}[t!]
    \centering
\includegraphics[width=10 cm]{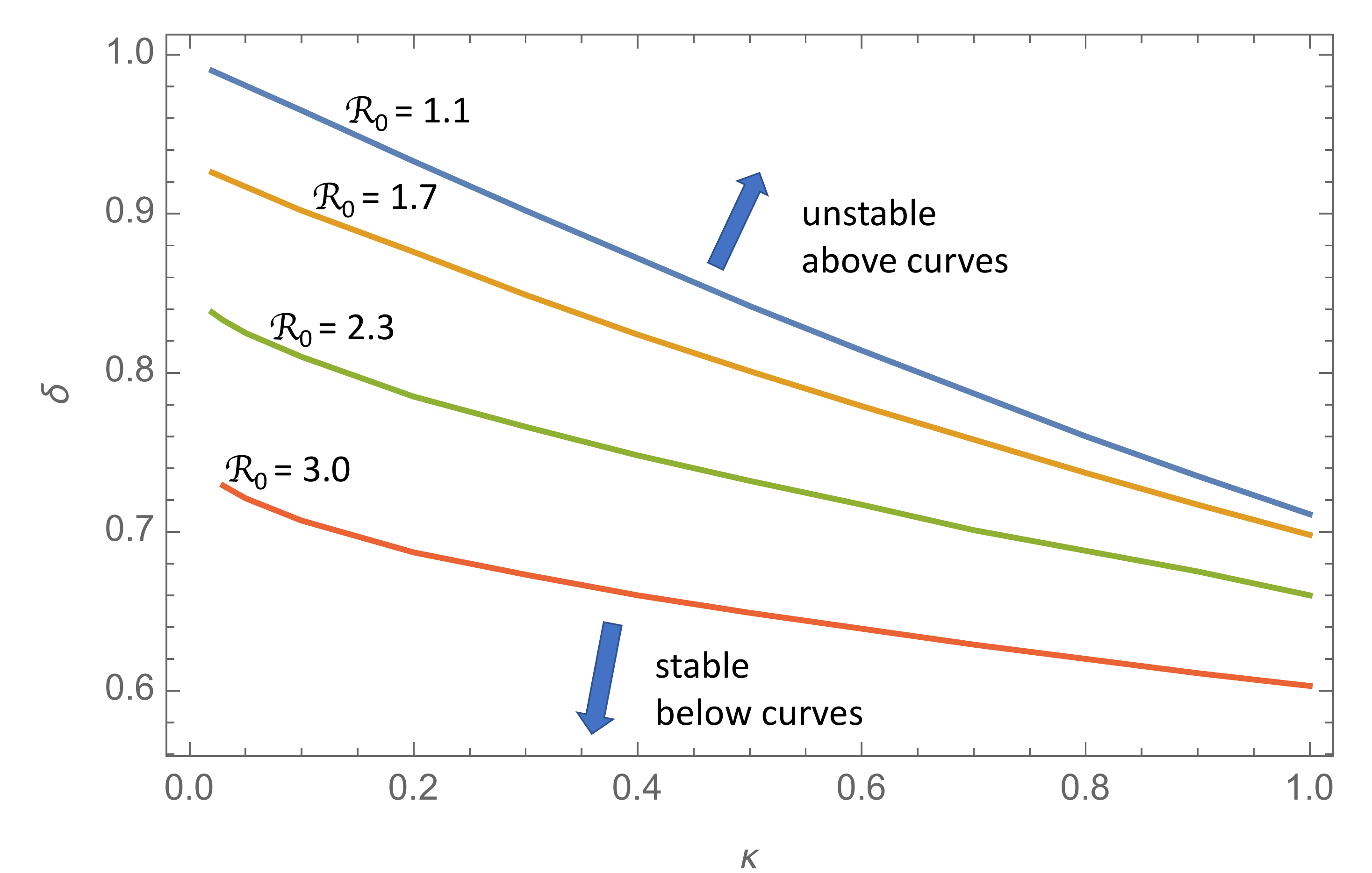}
    \caption{\footnotesize The figure depicts the transition between stable, decaying oscillations, like those shown in Figure \ref{fig:1} (c) and (d), and  unstable, growing oscillations, as shown in Figure \ref{fig:1} (e) and (f). The curves represent the parameters $\kappa$ and $\delta$ at the transition points, for four values of ${\cal R}_0$. At these transition points, the solution is a limit cycle like the one shown in Figure \ref{fig:1} (a) and (b). For each ${\cal R}_0$, the unstable region of the $(\kappa,\delta)$-space is the region above the corresponding curve.}
    \label{fig:2}
\end{figure}

Figure \ref{fig:2} shows isolines (curves of constant ${\cal R}_0$) in the $(\kappa,\delta)$-plane for ${\cal R}_0=1.1,1.7, 2.3, 3.0$. For any value of  ${\cal R}_0$ in this range and $\kappa$ in the range $(0.2,1.0)$, the solution is a growing oscillation if the $(\kappa, \delta)$-point is located above the isoline in the ($\kappa,\delta$)-plane, and a damped oscillation below this line. 

The take-home message from this figure is that there is a transition from a series of damped epidemic waves to a series of growing waves as the response delay $\delta$ exceeds a critical transition threshold $\sim 1$, or in dimensional units, $\sim 8$ days. Thus, this simple model suggests that a policy responding blindly to the actual incidence rate delayed by more than approximately a week may lead to a succession of epidemic waves of increasing amplitude.

\section{Discussion}\label{sec:discuss}

%\begin{listing}[H]
%\caption{Title of the listing}
%\rule{\textwidth}{1pt}
%\raggedright Text of the listing. In font size footnotesize, small, or normalsize. Preferred format: left aligned and single spaced. Preferred border format: top border line and bottom border line.
%\rule{\textwidth}{1pt}
%\end{listing}

Our model system, Eqs. (\ref{5}) and (\ref{6}), is an extremely simplified representation of a complex reality, although Eq. (\ref{6}) is probably far easier to accept than Eq. (\ref{5}), since it just describes the balance between new infections produced in a population where the instantaneous reproduction number is ${\cal R}(t)$ and the reduction of the number of infections due to recovery or death. Eq. (\ref{5}), on the other hand, aspires to encapsulate the  complex social dynamics that determines the evolution of ${\cal R}(t)$ in one single delay differential equation. Such a simple representation of a complex reality is certainly wrong, but may still be a valuable supplement to purely qualitative reasoning over the socio-political process that determines the response to a changing disease burden. 

The potentially disastrous effect of delayed NPI response was recognized by Pei, Shandula, and Shaman (2020), employing a metapopulation transmission model  and data on infections, deaths and human mobility in the United States \cite{Pei}. Their findings indicate that if control measures and reductions of ${\cal R}(t)$ had been implemented just 1 to 2 weeks earlier, substantial cases and deaths could have been avoided.  It is concluded that rapid detection of increasing case numbers and fast reimplementation of control measures are needed to control repeated outbreaks. They run a considerably more complex transmission model than our Eq. (\ref{6}), but make no attempt to model the social NPI response. They rather project the disease spread under factual and counterfactual NPI scenarios, and hence, the possibility that NPI delays may lead to a succession of epidemic waves of increasing amplitude is not explored.

The idea of NPIs trigged as incidence rates exceed a certain threshold  is not new either. In a paper on strategies for mitigation and suppression of COVID-19 in countries of different income level, Walker et. al (2020) \cite{Walker} modeled an oscillatory pattern of occupancy in intensive care units (ICUs) by assuming NPIs resulting in instantaneous reduction of 75\% in ${\cal R}$ from a basic level  3.0 each time the threshold is exceeded, and a duration of the NPIs of 1 month. Thus, in that model ${\cal R}(t)$ flips between 0.75 and a value that starts at 3.0 but is slightly reduced in successive flips as herd immunity starts to emerge. By construction, this model will not allow the oscillations of disease incidence rate to grow in amplitude;  the incidence threshold cannot be exceeded since the reproduction number drops instantaneously below 1 once the threshold is attained. The model presented here differs from \cite{Walker} in two important respects: 
\begin{enumerate}
\item The effect on ${\cal R}(t)$ of crossing the incidence threshold is not an instantaneous shift, but an effect on the rate of change $d_t \ln{\cal R}(t)$ which is proportional to the deviation from the threshold. In this way one allows for an inertia in the response, which gives rise to a damped oscillation around the threshold incidence. This can be seen from introducing the perturbations $\tilde{\cal R}(t)={\cal R}(t)-1$, $\tilde{I}(t)=I(t)-1$ around the equilibrium $({\cal R}, I)=(1,1)$, and linearizing Eqs. (\ref{5}) and (\ref{6}) for $\delta=0$, which yields the damped harmonic oscillator equation;
\begin{linenomath*}
\begin{equation}
d_t^2\tilde{I}+\kappa d_t\tilde{I}+\kappa \tilde{I}=0,
\end{equation}
where the ``friction  coefficient" and the ``spring constant" both are the same and given by $\kappa$.
\end{linenomath*}
\item We allow for a delay $\delta$ in the NPI-response with respect to the time the incidence threshold is crossed, and demonstrate that a sufficiently large delay may lead to a transition from a damped oscillation to a growing oscillation.
\end{enumerate}
The only damping taking place in \cite{Walker} is a very slow reduction in effective reproduction number arising from emerging herd immunity, i.e., reduction of the fraction of susceptible individuals, $S$. This damping effect is not taken into account in our model, since it assumed to be a negligible effect until vaccines effective against disease transmission have been rolled out in all adult age groups.

The absence in \cite{Walker} of delay in the social response and the absence of a reduction in the response strength due to intervention fatigue (as explored in \cite{Rypdal2020}), also preclude the possibility of growing oscillations. Thus, the value of the scenarios depicted in \cite{Walker} is primarily to quantify the fraction of the time societies will have to spend in lockdown in order to keep ICU occupancy below a given threshold, under the rather unrealistic assumption of full societal control of the reproduction number.

Under the assumption of an infectious period of $\alpha^{-1}=8$ days, and reasonable values of the parameters $\kappa$ and ${\cal R}_0$, our model predicts that the transition from damped to growing oscillation occurs for a response delay $t_d$ of the order of one week. This number should of course be taken with a grain of salt, considering the simplicity of the response model. Nevertheless, it is disturbing that this critical response time turns out to be so short that it may be impossible, even under ideal circumstances, to avoid this regime of recurrent waves of growing amplitudes, provided we assume that NPIs are implemented  by only using information about the most recent available case notification rate.

A valid objection to this assumption, and hence to Eq. (\ref{5}), is that governments have access to far more information when they lift on existing NPIs and decide on new ones.  They can consider existing trends in the incidence rates, not just the the most recent reported recordings, and model projections are available. On the other hand, it is clear for anyone who follow the public debate on the necessity of interventions, that it is extremely difficult for policy makers to gain public acceptance for precautionary interventions, even in cases  when a pattern has been repeated several times in the past. The common belief that the present wave is the last one seems to be a major hurdle to successful control  of the epidemic.

The tendency of repeated and stronger  epidemic waves  has been ubiquitous across the world's countries as we move into the second year of the pandemic, and a major part of the explanation may be the lack of ability of governments to take precautionary action. The key message from the present work is that relatively small changes in governments' ability to respond  in a more precautionary manner may have profound effects on controlling and mitigating new outbreaks. The main challenge seems to be convey this insight to policy makers, the media, and others who shape the public opinion.

\end{document}